\documentclass[12pt,a4paper]{article}

\usepackage{amsmath,amssymb,amsfonts,amsthm,latexsym,amscd}
\usepackage[english]{babel}
\usepackage{graphics}
\usepackage{framed}
\usepackage[mathscr]{eucal}

\usepackage[all,2cell]{xy} \UseAllTwocells 		

\newcommand{\C}{\mathcal{C}}

\newcommand{\G}{\mathcal{G}}
\renewcommand{\H}{\mathcal{H}} 	
 
\newcommand{\bbc}{{\mathbb{C}}}

\DeclareSymbolFont{rsfs}{U}{rsfs}{m}{n}
\DeclareSymbolFontAlphabet{\scr}{rsfs}

\newcommand{\Ef}{\scr{E}}

\DeclareMathOperator{\Ob}{Ob}

\DeclareMathOperator{\Homeo}{Homeo}

\DeclareMathOperator{\ls}{span}

\newtheorem{theorem}{Theorem}[section]				
\newtheorem{proposition}[theorem]{Proposition}

\theoremstyle{definition}        
\newtheorem{definition}[theorem]{Definition}

 \newtheorem{example}[theorem]{Example} 
  \newtheorem{examples}[theorem]{Examples}

\title{Applications to quantum gravity from C*-bundles}

\author{\normalsize  
Rachel A.D. Martins 
\\
\normalsize  \textit{Centro de An\'alise Mathem\'atica e Sistemas Din\^amicos, Departamento de
Mathem\'atica,}
\\
\normalsize \textit{Instituto Superior T\'ecnico, Universidade T\'ecnica de Lisboa,}
\\
\normalsize \textit{Av.~Rovisco Pais 1, 1049-001 Lisboa, Portugal}
 \thanks{Research supported by Funda\c{c}$\tilde{\mathrm{a}}$o   para
as Ci\^encias e a Tecnologia (FCT)
including programs POCI 2010/FEDER and SFRH/BPD/32331/2006. Email: rmartins-at-math.ist.utl.pt}}

\begin{document}

\maketitle

\begin{abstract}
 Applications to quantum gravity of some results in C*-algebras are developed. We open by describing
why algebra may be an integral aspect of quantum gravity. By
interpreting the inner automorphisms of a C*-algebra as families of parallel transports on a C*-bundle,
we define a notion of generalised connection for C*-bundles, implying that C*-bundle dynamical systems
can be used to study a Dirac operator's spectrum and also the generator
of the modular group. The fact that their dynamics preserves the bundle structure, means that C*-bundle
dynamical systems are relativistic systems. We describe a 2-category to unify Einstein's
two uses of the phrase general covariance, give concrete examples and construct a generally covariant
quantum system. Motivations include the point of view of virtual points hiding in fibres of
a C*-bundle over $X$ and that physical geometry is instrinsically non-local.
 \end{abstract}

\section{Introduction: Generalised spaces}

Algebra can be thought of as a laboratory, especially as gravity is a
branch of geometry and geometry can be viewed through the window of algebra. The interpretations and
applications of some previous results appear only in the form of passing comments in papers designated
by operator algebra, so even those brief comments are unlikely to reach physicists working in quantum
gravity who might be interested in them. In these notes we unify and develop those applications. We
try to distinguish clearly between interpretational results and discussion, which is kept to a minimum.
The algebraic technicalities are purposefully kept to a minimum.
\\

Descartes, Grothendieck and
Ehresmann each viewed geometry through the window of algebra. Ehresmann said that geometry is the study
of differentiable categories (of which Lie groupoids are the most basic example but in the context of
Quillen and Connes cyclic cohomology, a Morita-Rieffel category of imprimitivity bimodules can be
viewed as a generalisation) and Grothendieck replaced the traditional continuous topologies with
generalised topologies without points. Also from very early on, Heisenberg was asking for non-commuting
configuration coordinates $[q_i,q_j] \neq 0$. Crane advocates that categories are an
indispensable ingredient in quantum gravity. Of course, category theory is a topic in
algebra so if quantum gravity is fundamentally categorical, then quantum gravity is fundamentally
algebraic and so would benefit from more algebraic concepts. The distinguished direction of
time often leads to theories breaking covariance but Connes and Rovelli made a
physical interpretation of a distinguished state that already existed naturally in the intrinsic
dynamical properties of a Von Neumann algebra \cite{thermal}, and that indicated the emergence of a
direction of time (pointing in the direction of increasing entropy). The path-integral formulation of
quantum
gravity involves various approaches to discretisation of space-time through choices of triangulations of
the (traditionally smooth) space-time manifold and in \cite{mcqg} Crane explains that through algebraic
methods the need to make that choice can be removed due to an a priori modelling of the quantum nature
of the space-time. The spectral action principle \cite{sap} by Connes and
Chamseddine with important contributions from Sch\"ucker \cite{forces,ncg and sm} involves a purely
algebraic equivalence principle on a non-commutative Kaluza-Klein-like internal space. In this model
the action is fully diffeomorphism invariant, albeit a classical action, this is another clear potential
benefit of involving algebraic methods in quantum gravity.
\\

Since quantum gravity involves trying to understand the nature of space-time, it will
necessarily involve space-time on the smallest scales, or the Planck length
scale. (Even though space-time-gravitational components of theories will not be tested directly on
that level due to the negligible effect of the
force, to understand gravity we first must understand more about what physical geometry
really is.) It is very clear that Analytical calculus provided one of the greatest ever inventions but
even so, Crane points out that it breaks down in the context of quantum space-time at the Planck length
scale since it relies
on infinitesimal distances as part of the structure of this powerful mathematical tool.
Fortunately there are already algebraic generalisations of
calculus such as Connes' and Quillen's cyclic cohomology which draws heavily on Morita-Rieffel theory,
which in itself has a rich geometrical significance.
\\

Although it is traditional to pass to a
non-commutative generalisation of a topological space by invoking the Gelfand-Naimark theorem and then
viewing all non-commutative algebras as generalised algebras of functions, it is also true
that $C_0(X)$ is the algebra of sections of a line bundle over $X$, so there is an argument to think of
non-commutative algebras as generalised algebras of sections, where \emph{the fuzzy points hide in the
fibres} over (a tangible) $X$ instead of in a fuzzy space ``$X$''. The following candidate for a quantum
space-time is not finished (although it was described in \cite{sc}) but it has to be mentioned in order
to explain the context of these notes. 

Let $E^0$ be a C*-bundle whose
fibres are given by simple matrix algebras of varying dimension so that the enveloping algebra
$C^*(E^0)$ is given by $A= \bigoplus M_{n_i}(\bbc)$ $i=1...m$ or alternatively let $A$ be an
approximately finite algebra with direct limit $A$. The fibres $A_i$ are all members of the same Morita
equivalence class. Now consider the linking algebra $B$ that $A$ is embedded in, coming from the
Morita equivalence bimodules (for instance, $\bbc^2$ is a Morita equivalence bimodule over $M_2(\bbc)$
and $\bbc$). Just to illustrate for readers unfamiliar with this context, a typical element of a
linking algebra, 

\begin{equation}
e = \left(   \begin{array}{cc}
e_{\gamma \gamma^{\ast}} &    e_{\gamma}     
\\
e_{\gamma^{\ast}}       &     e_{\gamma^{\ast} \gamma}
\end{array}  \right)
\end{equation}

where $e_{\gamma \gamma^{\ast}} \in A_1$, $e_{\gamma^{\ast} \gamma} \in A_2$, $e_{\gamma} \in M_{AB}$ an
$A$-$B$-bimodule and $e_{\gamma^{\ast}} \in M_{BA}$ a $B$-$A$-bimodule. \cite{Rieffel Morita}.

 Every space-time needs a tangent bundle, and this will come from the linear span of derivations from
$A$ into $B$: using basic category theory, one may define a representable sheaf from such a
category of Morita equivalence bimodules to generalise the sheaf of sections of the tangent bundle.
Further work is to check to what extent this satisfies the criteria of a Crane quantum space-time
\cite{mcqg,wqg} and
how closely it satisfies the axioms of an Algebraic quantum field theory and also how it relates to
Rovelli's relativistic quantum mechanics.
\\

In \cite{group of loops} following Barrett and Anandan, Lewandowski fully
constructed a vector bundle together with a connection just from the data afforded by the holonomy
group. Equivalently, a representation of a Lie groupoid on a vector bundle  $\Ef$ is equivalent to a
choice of connection on $\Ef$. These statements imply that geometry can be fully described in a
non-local way. Let $M$ be a simply connected manifold with tangent bundle $TM$. When an integration of a
tangent vector is performed, these two things happen: an arrow in a differentiable category is obtained
(i.e. the groupoid $M \times M$) and the algebra $C^*(TM)$ is deformed to a non-commutative algebra of
observables $C^*(M \times M)$. This implies there is an intimate connection between category theory and
non-commutative geometry, afterall, they are each topics from both geometry and algebra.

\section{Generalised connections and geodesics}

In short, a C*-bundle is a bundle of C*-algebras over a topological space where the enveloping algebra
is a C*-algebra $C^*(E^0)$ or $A$. For a full definition see \cite{Dixmier} or \cite{Fell Doran}.
For Fell, Dixmier and many others, a C*-bundle $(E^0,\pi,X)$ is not equivalent to a fibre bundle with
additional structures. In particular, it may be non-locally trivial. Also recall that the general linear
groupoid $GL(\Ef)$ of a vector bundle $\Ef$ is the groupoid of all isomorphisms between pairs of fibres.
Let $GL(E^0)$ denote the groupoid of isometric *-isomorphisms between pairs of fibres of $E^0$. Let
$A$ be represented on a separable Hilbert space $\H$. 

Some simple examples of C*-bundles:

\begin{examples}
 \begin{enumerate}
  \item Let $A = \bigoplus_i M_{m_i}(\bbc)$ so fibres of $E^0$, denoted
$E^0_x$, $x \in X$, also denoted $A_x$, are simple matrix algebra summands $M_m(\bbc)$ of varying
dimension $m$. 
\item As the previous example except that all fibres are of equal dimension $m$. 
\item Let $E^0$ be a 1-dimensional C*-bundle over a space $X$. 
 \end{enumerate}
\end{examples}

Recently, (reversible) C*-bundle dynamical systems were defined:
\begin{definition} \cite{DS}
 A \emph{reversible C*-bundle dynamical system} $(E^0,A,\G_{\sigma})$ is given by a C*-bundle $(E^0,A)$
and a 1-parameter covariance group $\G_{\sigma}$ of C*-bundle inner automorphisms.
\end{definition}

where the elements $f$ of $\G_{\sigma}$ are of the form $f(a) = U_t a U_t$ with $U_t = e^{i \sigma t}$
for a possibly unbounded self-adjoint operator $\sigma$ and with $a \in A$. Now let $a \in E_x^0$. Each
$f$ restricts to a family of fibrewise mappings $\alpha:E_x^0 \to E_y^0$, $\alpha_x(a)=uau^*$
which from now on we call parallel transports.

In \cite{DS} the interpretation was given to $\sigma$ as an operator whose eigenvalues coincide with
those of a Dirac operator $D$ for a spectral triple $(A,\H,D)$ with $A$ and $\H$ as above. Next we
develop this as an application to quantum gravity with algebra. 
\\

In the special case that $E^0$ is a vector bundle ($\H$ may be infinite dimensional if desired) then
each $U_t$ induces fibrewise isomorphisms or parallel transports $\alpha: E^0_x \to E^0_y$,
$\alpha(a) = uau^*$ with $u=e^{i \sigma_x t}$ where from the set of components $\sigma_x \in A_x$
extended over all of $X$, one can infer the connection on the bundle. This leads to a notion of
generalised connection:

\begin{definition}
 Let $(E^0,\G_{\sigma})$ be a reversible C*-bundle dynamical system. Then a \emph{generalised
connection} is given by the operator $\sigma$.
\end{definition}

In Connes' non-commutative counterpart of Einstein's equivalence principle, he constructs a curved
space Dirac operator on a Riemannian spin manifold from an
intial flat Dirac operator beginning from the data of a generalised diffeomorphism, that is, an inner
automorphism of the spectral triple algebra. His fluctuations formula:

\begin{equation} \label{fluctuations}
  D^f = \sum_{\textrm{finite}} r_j L (\sigma_j) D L(\sigma_j)^{-1}, ~~r_j \in \mathbb{R}, ~~ \sigma_j
\in \textrm{Aut}(A)
\end{equation}

where $L$ is the double valued lift of the automorphism group to the spinors. For the calculation of
the spectral standard model action see \cite{sap}. The result is the general form of the Dirac
operator with arbitrary curvature and torsion:

\begin{equation} \label{general D}
 D =  \sum_i c_i \Big(\frac{\partial}{\partial x}_i + \omega_i \Big)
\end{equation}

We infer that much if not all of the geometrical data needed for the connection on the spinor bundle can
be constructed from the eigenvalues of $D$. Therefore, and especially in finite dimensions, the
interpretation of $\sigma$ as a certain kind of ``generalised connection'' (the kind defined precisely
above) is arguably sensible.
\\

In \cite{spectral C} non-commutative geodesics were defined. For a C*-bundle with isomorphic fibres,
this notion of geodesic is given by the components $u=e^{i \sigma_x t}$ of each $U_t$. In
\cite{essay}
Connes first mentioned non-commutative geodesics as paths $\gamma = e^{i \vert D \vert t}$, where $\vert
D \vert$ refers to eigenvalues of $D$. With the tangent sheaf replacing the ``deformed tangent
bundle'' $M \times M$, then
$u \in B$ or $\gamma$ is exactly a path in the quantum space formalised by $E^0$. 
\\

 (A full treatment of the case of example
2.1(1) where fibres are not isomorphic will require generalising to irreversible C*-bundle dynamical
systems with 1-parameter semigroup of *-endomorphisms implemented by partial isometries.) 
\\

In the context where the isotropy groups of the category $GL(E^0)$ are Lie groups, then $\sigma_x$ will
be an element of the Lie algebra of the group at $x$. (In order to study a space of generalised
connections $\{ \sigma \}$ a deeper study of the dynamical
properties of Von Neumann algebras and modular groups might be needed, especially following
ideas of Bertozzini including in \cite{BCL mst}. Then it might be possible to
calculate states for quantum gravity using algebraic methods.)
\\

In \cite{DS} it was shown that the following formula characterises a reversible C*-bundle dynamical
system:

\begin{equation} \label{partition}
\Phi_{\hookrightarrow} = \sum_{m=1}^n \prod_{i=1}^m U_{g_i}, ~~~~g_i \in \G_{\sigma},~~ i=1..m,
~~m=1..n. 
\end{equation}

$\Phi_{\hookrightarrow}$ resembles a partition function since it is a sum over geometrical states and
the product consists of parallel transports. (More work is needed to make this precise but the
fundamental excitations $u_{g^*g \in X \times X} \in A_x$ should correspond to the components
$\sigma_x$.)

\section{General covariance 2-categories}

A C*-bundle dynamical system satisfies an additional condition called the covariance condition.  This is
the property that the inner automorphisms $f$ respect the bundle structure. For
dynamical systems it means that a physical configuration space is singled out (and other mathematical
consequences that we treated in \cite{DS}). Let $G$ denote the pair groupoid $X \times X$ and let $\G$
denote the group of bisections of $G$, which is identified with the group of homeomorphisms
$f_0 \in \Homeo(X)$. The physical significance of the covariance condition is that in given physical
theory, if $X$ is a space-time and if $\G=\Homeo(X)$ is replaced by the symmetry
group of the theory (perhaps adding smoothness conditions) then the theory is covariant with respect to
the group $\G$. So the inner automorphisms of a C*-bundle covariant with respect to all coordinate
transformations or diffeomorphisms will give rise to a generally covariant theory whereas if the
symmetry group $\G$ of the theory is the Poincar\'e group then
the theory will be Poincar\'e covariant. Since C*-dynamical systems are quantum mechanical
systems, this means that a C*-bundle dynamical system is a relativistic quantum system. 

\begin{definition}
 C*-bundle inner automorphisms must preserve the bundle structure. The corresponding condition,
 \begin{equation}
  \pi \circ f = f_0 \circ \pi
 \end{equation}
 is called the \emph{covariance condition}.
\end{definition}

By \emph{general covariance}, Einstein meant two things: (i) the theory is diffeomorphism invariant,
the 
physics being independent of the choice of coordinates and he also meant that (ii) there was no given
initial
background geometrical data, the curvature of space-time originated from fluctuations away from the
initial flat metric. A 2-category can capture general
covariance where the first meaning (i) comes into the 1-morphisms and the second meaning (ii) is
captured by the 2-morphisms: 

By \emph{general covariance 2-category} we mean a 2-category $\C$ with an object space $\Ob(\C)$ with a
disjoint union operation to model a space-time, with 1-morphisms to model parallel transports in a
covariant way and 2-morphisms to model fluctuations of the gravitational field implementing the
equivalence principle. 

(Ultimately, a general covariance 2-category should include an
inverse category of partial maps so that bundles with non-isomorphic fibres can be included).

\begin{example} \label{example 1}
 Let $E^0$ be a C*-bundle with isomorphic fibres and let $GL(E^0)$ be the groupoid of all isometric
*-isomorphisms between pairs of fibres $(E^0_x,E^0_y)$. Let $A=C^*(E^0)$ be represented on a separable
Hilbert space. The following describes a general covariance 2-category $\C$.
\begin{itemize}
 \item Objects are the fibre C*-algebras $A_i$ of $E^0$. $\Ob$ has a direct sum
operation. 
 \item 1-arrows are all the elements of $GL(E^0)$, $a \mapsto uau^*$ with $u=e^{i s_x t}$ with $s_x$ the
component at $x$ of a possibly unbounded self-adjoint operator $s$, for each choice of $u$.
 \item 2-arrows are given by assignments $s_x \mapsto u s_x u^*$
\end{itemize}
\end{example}

\begin{example}[General covariance 2-group]
 The previous example gives rise to a 2-group such that 1-arrows are the global bisections of
$GL(E^0)$, which, exactly because they are bisections, satisfy $\pi \circ f = f_0 \circ \pi$, (the
mathematical reason for this is made clear in \cite{DS}.) 2-arrows are fluctuations: $s \mapsto U s
U^*$ to be compared with formula \ref{fluctuations}. Note that a 2-arrow is almost iterative of a
1-arrow.
\end{example}

\begin{example} \label{example 2}
 \begin{enumerate}
  \item Objects: the same as in the previous example.
  \item 1-arrows: Morita equivalence bimodules.
  \item 2-arrows: bimodule isomorphisms. 
 \end{enumerate}
\end{example}

Example \ref{example 2} is a coarse-grained version of example \ref{example 1} since:

each 1-arrow of \ref{example 2} is given by the linear span  of the
commutators $[s,c]$ over $c \in A_x \oplus A_y$ for any $s_x$ and is the $A_x$-$A_y$-bimodule
$\Omega^1_{x}$ which we also denote
$E_{xy}$. 

On the 2-arrow level, $\ls_{c \in A_x \oplus A_2} [u s_x u^*,c] = u E_{xy} u^*$ because for all $c' \in
A_x \oplus A_y$ there exists a $c$ such that $[u s u^*,c]=u[s,c']u^*$, and  $e \mapsto u e
u^*$ for all $e \in E_{xy}$ defines a bimodule isomorphism.


\begin{proposition}
 A (reversible) C*-bundle dynamical system together with a general covariance 2-group is a generally
covariant quantum mechanical system.
\end{proposition}
\begin{example}
 Any spectral C*-category (see \cite{spectral C} for the definition) with isomorphic objects and any
spectral triple of an equivalent description.
\end{example}
It remains to verify if these systems satisfy requirements asked for in the literature on and
related to covariant quantum mechanics.

\section{Inner categorification}

The only news in this section is just discussion.

The procedures in ``inner categorification''  of spectral triples \cite{spectral C} (resulting in
spectral C*-categories) came from the point of view on non-local geometry explained in the last
paragraph of the first section together with the view that fuzzy points might be hiding in the fibres
rather than in the base manifold, reflecting the bundle-like nature of a Riemannian spin manifold.
Connes explains that a curved Dirac operator only arises from formula  \ref{fluctuations} in the
non-commutative case, so it was suggested in \cite{spectral C} that space-time curvature might arise
from non-commutativity of space-time. More studies involving $\sigma_x$ as a Lie algebra element might
shed light. 
In the case of a C*-bundle dynamical system with a line bundle $E^0$, the fibre is $\bbc$, the bundle
endomorphisms form a group and so the system is
reversible. In contrast, since not all fibres are given by $\bbc$, a dynamical system with $E^0$ from
example 2.1(1) will involve endomorphisms between fibres that are not invertible, which means that some
of the dynamics will be irreversible and the question arises if the time irreversibility we observe has
emerged from
the above curving effect of the non-commutativity in the space $E^0$ (especially if the dynamical system
is based on a
Von Neumann crossed product \cite{thermal}).

With inner categorification, together with the details of spectral triple morphisms appearing in
 \cite{BCL Cncg}, \cite{BCL stm}, \cite{BCL cqp} and other works by those authors and others, together
with a bundle dynamical perspective, it should be possible to construct an algebraic counterpart to
TQFTs in a canonical way. In particular, these C*-categories are concrete, which means they already
possess a representation or functor into a category of Hilbert spaces and bounded linear maps, Hilb.

\section{Acknowledgements}
Many thanks to John Barrett, Paolo Bertozzini, Louis Crane and Pedro Resende for the discussions
(alphabetical order). See title page for affiliation.


\begin{thebibliography}{999}


\bibitem[BCL1]{BCL stm}
P. Bertozzini, R. Conti, W. Lewkeeratiyutkul  (2006).
A category of spectral triples and discrete groups with length function
\textit{Osaka J. Math.} Volume 43, Number 2 (2006), 327-350. 


\bibitem[BCL2]{BCL mst}
P. Bertozzini, R. Conti, W. Lewkeeratiyutkul 
Modular Theory, Non-Commutative Geometry and Quantum Gravity,
\textit{Special Issue ``Noncommutative Spaces and Fields'', SIGMA} 6:067, (2010) 
Arxiv: 1007.4094

\bibitem[BCL3]{BCL cqp} 
P. Bertozzini, R. Conti, W. Lewkeeratiyutkul  
\textit{Non-commutative geometry, categories and quantum physics.} 
Arxiv: 0801.2826 (2007).

\bibitem[BCL4]{BCL Cncg}
P. Bertozzini, R. Conti, W. Lewkeeratiyutkul  
Categorical non-commutative geometry,
\textit{J. Phys.: Conf. Ser.} 346 012003 (2012)

\bibitem[CC]{sap} 
A.H. Chamseddine, A. Connes (1997)
The spectral action principle.
\emph{Comm. Math. Phys.} Vol.186 (1997), N.3, 731-750.

\bibitem[C]{essay} A. Connes, Noncommutative geometry and physics,
http://alainconnes.org/docs/einsymp.pdf

\bibitem[CR]{thermal} A. Connes, C. Rovelli, 
Von Neumann algebra automorphisms and time-thermodynamics relation in general covariant quantum
theories, 
\textit{Class. Quant. Grav. 11: 2899-2918 (1994)}
Arxiv: gr-qc/9406019.

\bibitem[LC1]{mcqg}
L. Crane,
Model categories and quantum gravity,
Arxiv: 0810.4492 (2008).

\bibitem[LC2]{wqg}
L. Crane,
What is the mathematical structure of quantum spacetime?
ArXiv:0706.4452v1 (2007).

\bibitem[D]{Dixmier} 
Dixmier J (1982). 
$C^*$-algebras, 
North-Holland Publishing company, English translation

\bibitem[FD]{Fell Doran} 
J. Fell J, R. Doran (1998). 
Representations of $C^*$-algebras, Locally Compact Groups and Banach $*$-algebraic bundles,
vol~1-2, Academic Press

\bibitem[L]{group of loops}
Jerzy Lewandowski,
Group of loops, holonomy maps, path bundle and path connection,
\textit{Class. Quantum Grav. 10 (1993) 869-904}

\bibitem[M3]{DS}
R.A.D. Martins,
C*-bundle dynamical systems,
Arxiv: 1402.1206 (2014)

\bibitem[M1]{sc} 
R.A.D. Martins,  
Categorified noncommutative manifolds,
\textit{International Journal of Modern Physics A}, 15 (2009).
Arxiv: math.ph/0811.1485 

\bibitem[M2]{spectral C}
R.A.D. Martins,
Spectral C*-categories and Fell bundles with path-lifting.
Arxiv: 1308.5247 (2013).


\bibitem[R]{Rieffel Morita}
M. A. Rieffel (1982). 
Morita equivalence for operator algebras,
\textit{Proceedings of Symposia in Pure Mathematics (American Mathematical Society)} 
38: 176-257.


\bibitem[S1]{forces} 
T. Sch\"ucker, 
Forces from Connes' geometry, 
ArXiv:hep-th/0111236.
\textit{Lect. Notes Phys}. 659:285-350 (2005).


\bibitem[S2]{ncg and sm} 
T. Sch\"ucker, 
Noncommutative geometry and the standard model, 
Arxiv: hep-th/0409077. 
\textit{Int. J. Mod. Phys}. A20:2471-2480 (2005).


\end{thebibliography}
\end{document}